\documentclass[apj]{emulateapj}
\usepackage[utf8]{inputenc}
\usepackage{graphicx}
\usepackage{amsmath}
\usepackage{natbib}

\begin{document}

\title{Towards a direct measure of the Galactic acceleration}

\author{Sukanya Chakrabarti\altaffilmark{1,12}, 
Jason Wright \altaffilmark{2}, Philip Chang \altaffilmark{3}, Alice Quillen \altaffilmark{4}, Peter Craig \altaffilmark{1},
Joey Territo \altaffilmark{1},
Elena D'Onghia \altaffilmark{5}, Kathryn V. Johnston \altaffilmark{6,11}, Robert J. De Rosa \altaffilmark{7}, Daniel Huber \altaffilmark{8}, Katherine L. Rhode \altaffilmark{9}, Eric Nielsen \altaffilmark{10}}

\altaffiltext{1}
{School of Physics and Astronomy, Rochester Institute of Technology, 84 Lomb Memorial Drive, Rochester, NY 14623; chakrabarti@astro.rit.edu}
\altaffiltext{2}
{Department of Astronomy \& Astrophysics and \\
Center for Exoplanets and Habitable Worlds and \\ 
Penn State Extraterrestrial Intelligence Center \\
525 Davey Laboratory \\ The Pennsylvania State University \\ University Park, PA, 16802, USA}
\altaffiltext{3}
{Department of Physics, University of Wisconsin-Milwaukee, 3135 North Maryland Avenue, Milwaukee, WI 53211}
\altaffiltext{4}
{Department of Physics and Astronomy, University of Rochester}
\altaffiltext{5}
{Department of Astronomy, University of Wisconsin-Madison}
\altaffiltext{6}{Department of Astronomy, Columbia University, New York, NY}
\altaffiltext{7}{European Southern Observatory, Alonso de C\'{o}rdova 3107, Vitacura, Santiago, Chile}
\altaffiltext{8}{University of Hawaii}
\altaffiltext{9}{Department of Astronomy, Indiana University, Bloomington, IN 47405}
\altaffiltext{10}{Stanford University}
\altaffiltext{11}{Center for Computational Astrophysics, Flatiron Institute, New York, NY}
\altaffiltext{12}{Institute of Advanced Study, 1 Einstein Drive
Princeton, New Jersey
08540 USA}
\begin{abstract}

High precision spectrographs can enable not only the discovery of exoplanets, but can also provide a fundamental measurement in Galactic dynamics.  Over about ten year baselines, the expected change in the line-of-sight velocity due to the Galaxy's gravitational field for stars at $\sim$ kpc scale distances above the Galactic mid-plane is $\sim$ few - 10 cm/s, and may be detectable by the current generation of high precision spectrographs.  Here, we provide theoretical expectations for this measurement based on both static models of the Milky Way and isolated Milky Way simulations, as well from controlled dynamical simulations of the Milky Way interacting with dwarf galaxies.  We simulate a population synthesis model to analyze the contribution of planets and binaries to the Galactic acceleration signal.  We find that while low-mass, long-period planetary companions are a contaminant to the Galactic acceleration signal, their contribution is very small. Our analysis of $\sim$ ten years of data from the LCES HIRES/Keck precision radial velocity (RV) survey shows that slopes of the RV curves of standard RV stars agree with expectations of the local Galactic acceleration near the Sun within the errors, and that the error in the slope scales inversely as the square root of the number of observations.  Thus, we demonstrate that a survey of stars with low intrinsic stellar jitter at kpc distances above the Galactic mid-plane for realistic sample sizes can enable a direct determination of the dark matter density.

\end{abstract}

\section{Introduction}
High-precision spectrographs have recently come online that are designed to search for Earth-sized planets orbiting Sun-like stars and are expected to have an instrumental precision of order 10 cm/s \citep{Pepe2010,Fischer2016, WrightRobertson}.The NEID spectrograph has been designed to have an instrumental precision of 30 cm/s, and was deployed last year on the WIYN 3.5m telescope \citep{NEID_optical}.  VLT's ESPRESSO has demonstrated precision better than 30 cm/s for quiet stars (Cabral et al. 2019).  Additionally, next generation Doppler spectrographs on 10m+ telescopes such as the Keck Planet Finder \citep{Gibson2016} will push such exquisite RV precision to fainter stars and larger distances.  These spectrographs can also be used to directly measure the Galactic acceleration, which gives a constraint on the dark matter density.  As such, it is likely the most fundamental measurement that can be made in Galactic dynamics.

To determine the nature of the dark matter particle from direct dark matter detection experiments requires an independent measure of the local dark matter density \citep{Read2014}, as extrapolated to the lab.  The traditional method is to estimate the acceleration using stellar kinematics, and accounting for the baryon budget in the solar neighborhood, infer the total density from the Jeans analysis (which assumes equilibrium) using stellar velocity dispersions \citep{Kuijken_Gilmore1989, HolmbergFlynn,McKee2015, Widmark_Monari2019}.  Other kinematical estimates besides the Jeans analysis have also been explored \citep{Salucci2010}.  In regions far from the Galactic mid-plane where baryonic processes do not affect halo shapes \citep{Prada2019}, the dark matter density may constrain the shape of the Milky Way's dark matter halo \citep{Read2014}, and how it is affected by cosmological accretion and satellite interactions via a study of its sub-structure.  This work is largely focused on measurements at a few kpc distances above the Galactic mid-plane.  A measurement of the Galactic acceleration would allow us to determine the viability of different dark matter models, as well as others including modified Newtonian gravity \citep{Milgrom1983,Milgrom2010}, and towards understanding the history of cosmic accretion in the Milky Way.

Recent work has shown that there are significant differences between the \emph{true} density measured in a simulation and the density inferred from using the Jeans approximation, especially in regions where the Galaxy is perturbed \citep{Haines2019}.  Analysis of Gaia DR-2 data has revealed the so-called phase-space spiral (Antoja et al. 2018), as well as the Enceladus merger (Helmi et al. 2018), clearly indicating a Galaxy that is out of equilibrium.

Earlier work \citep{Quercellini2008, Silverwood2019,Ravi2019} estimates that over a baseline of about ten years, high precision spectrographs should be able to directly measure the local acceleration of the Galaxy, i.e., relative to the solar acceleration.  The advent of high resolution spectrographs, coupled with a better understanding of stellar jitter \citep{Yu2018, Luhn2020}, now renders this fundamental measurement feasible.   Gaia's proper motion data will not have sufficient accuracy to detect the Galactic acceleration, even over ten year baselines \citep{Silverwood2019}.  Here, we show that targeted surveys with high precision spectrographs are a promising route that can deliver this measurement.

In this paper, we go beyond earlier work by laying out theoretical expectations from not only static models of the Milky Way, but also from dynamically evolving simulations of the Milky Way (\S 2) interacting with dwarf galaxies, and contrast the differences with static and isolated high resolution models of the Galaxy that include the effect of the Galactic bar.  In \S 3, we simulate a Galactic population composed of single stars, binaries, and planets, and thereby calculate the effect of contaminants (stellar binaries and planets) on the Galactic acceleration.   We analyze existing ten year data from the LCES HIRES/Keck precision radial velocity exoplanet survey of stars near the Sun for standard RV stars.  We show that the errors in the slopes of the RV curves, although non-Gaussian (arising from stellar jitter and instrumental noise), may be taken to be Gaussian for the purposes of quantifying the effect of this noise on the measured acceleration.  One of the significant advances of this work is that we show that (\S 3) the contamination to the Galactic signal from planets in a realistic sample size is very small.  We conclude in \S 5.

\section{Theoretical Expectations}
\subsection{Static \& isolated Milky Way models}

\begin{figure}[h]        
\begin{center}
\includegraphics[scale=0.55]{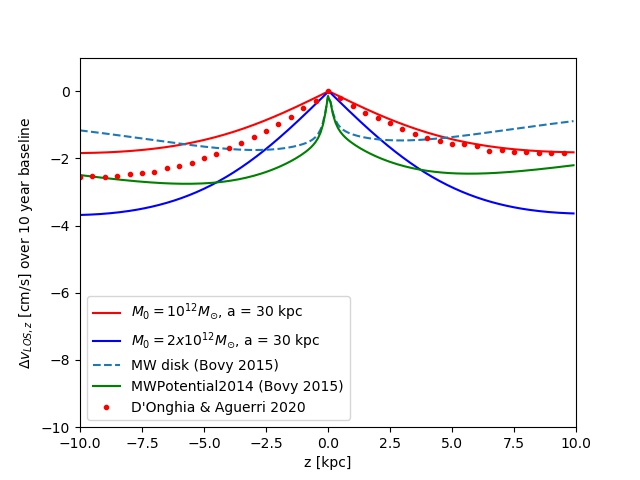}
\caption{Change in the line-of-sight velocity over a baseline of ten years in a Galaxy model with a dark matter halo having a Hernquist profile and a total mass of $2 \times 10^{12} M_{\odot}$ (blue line) and $10^{12} M_{\odot}$ (red line), both with scale length = 30 kpc.  Also shown is the contribution from Bovy 2015's Milky Way disk model (dashed blue line), and the MWPotential2014 potential (green line), and a recent high resolution isolated simulation of the Milky Way (D'Onghia \& Aguerri 2020; in red dots).  RV observations at $\sim$ kpc distances off the mid-plane near the Sun would produce measurable
changes in the line-of-sight velocity, and would primarily probe the dark matter component of the potential. \label{f:static}}
\end{center}
\end{figure}

For a baseline expectation for the local Galactic acceleration, we begin by considering static models of the Milky Way.  The radial acceleration is $\sim v_{c}^{2}/R_{\odot}$, where $v_{c}$ is the circular velocity of the Sun and $R_{\odot}$ is the Galactocentric radius of the Sun.  The GRAVITY Collaboration et al. (2018) has measured $R_{\odot}$ to high precision, and this measurement, combined with the tangential component of the solar peculiar motion, gives us $v_{c}$.  Thus, we have reasonably good bounds on the radial acceleration.  It will be useful to verify the direct acceleration method as compared to current constraints on the radial acceleration near the Sun; \cite{Silverwood2019} have discussed the range of radial accelerations for a specific MW potential model, which give $\sim$ few cm/s near the Sun.  Here, we focus on the vertical acceleration which is significantly more uncertain at $\sim$ kpc distances above the Galactic mid-plane.  

To illustrate the range of possible vertical accelerations, we consider a range for the Milky Way total mass that spans $\sim 1-2 \times 10^{12} M_{\odot}$, as found in the literature \citep{Watkins2019,Deason2019,PostiHelmi2019,Fritz2018,Piffl2014,BoylanKolchin2013}.  Figure \ref{f:static} depicts the change in the line-of-sight velocity in the vertical direction ($\Delta v_{\rm LOS,z}$) for stars in a Galaxy model that incorporates a dark matter halo having a Hernquist (1990) density profile, shown here for masses of $10^{12} M_{\odot}$ (solid red line) and $2 \times 10^{12} M_{\odot}$ (solid blue line), both with scale length of 30 kpc. Here, we have used the Galpy \footnote{http://github.com/jobovy/galpy} software \citep{Bovy2015}, to calculate the line-of-sight acceleration for a specified potential, relative to the Sun's acceleration.  We take the Sun's position to be at Galactocentric coordinates X=8.1 kpc, Y=0, Z = 0.05 pc.  The line of sight is taken to be from the Sun to some vertical height above or below the Sun.  The change in the line-of-sight velocity is shown here over a time baseline of ten years.  Also shown here is the contribution from the Milky Way disk model (dashed blue line) developed by \cite{Bovy2015}, and for the MWPotential2014 model for the total Milky Way potential as described in \cite{Bovy2015}.  $\Delta v_{\rm LOS,z}$ is negative and $\sim$ few - 10 cm/s for $z \geq$ 2 kpc off the Galactic mid-plane, at which point the potential is clearly dominated by the dark matter halo. 
  
Estimates of the acceleration using observations of the vertical kinematics and density of A and F stars in the context of the vertical Jeans equation \citep{HolmbergFlynn} at 1 kpc above the mid-plane give a $\Delta v_{LOS,z}$ of 1 cm/s, which is consistent with the range shown here.  

An important consideration for isolated models of the Milky Way is the effect of the bar on the Galaxy.  A recent high resolution $N-body$ simulation of the Milky Way \citep{DOnghiaAguerri2020} adopts a long-bar scenario (extending about 5 kpc from the Galactic center), which reproduces the formation of the Hercules stream, as well as other features in the Solar vicinity.  Figure \ref{f:static} also depicts a comparison to this simulation of the Milky Way (shown here in red dots).  Here, the Sun is placed at Galactocentric coordinates X= 7.15, Y = -3.8 kpc, following their work, and we compute the acceleration above and below the mid-plane from the Sun.  For this and the SPH simulations that we analyze below, we compute the acceleration directly from the particle information.  As is clear, the effect of spiral arms and the bar do not lead to a pronounced asymmetry in the vertical acceleration, and the magnitude of the acceleration is comparable to the static cases shown.

\subsection{Simulations of the Milky Way interacting with dwarf galaxies}

The Milky Way exhibits a plethora of signatures of interactions, including tidal streams such as that of the Sagittarius (Sgr) dwarf galaxy \citep{Ibata1994}, a warp and large planar disturbances in the HI disk \citep{LevineBlitzHeiles2006}, vertical waves in the stellar disk \citep{Xu2015}, as well as more recent discoveries from Gaia DR-2 including the Gaia Enceladus merger \citep{helmi2018}.  Thus, it is important to consider a dynamically evolving galaxy undergoing external perturbations, and its effect on the Galactic acceleration.

\begin{figure}[ht]        
\begin{center}
\includegraphics[scale=0.48]{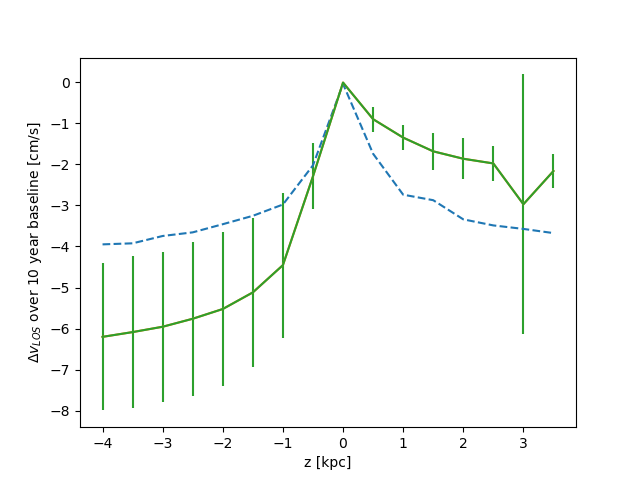}
\includegraphics[scale=0.48]{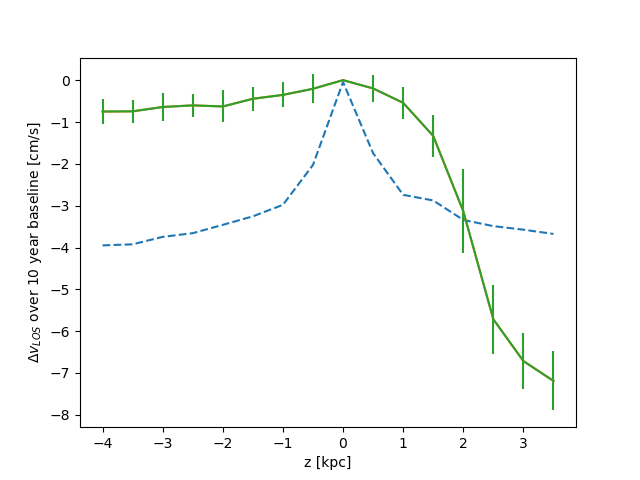}
\caption{(a) Change in the line-of-sight velocity for a simulation of Milky Way interacting with the Antlia 2 dwarf galaxy at present day (solid line) and at early times prior to the interaction with Antlia 2 (dashed line) \citep{chakrabarti2019}; solid line shows the average over a ring of radius r = 8 kpc, and the error bars show the standard deviation along the azimuth (b) $\Delta v_{LOS,z}$ for the interaction of the Sagittarius dwarf galaxy with the Milky Way \citep{chakrabarti2019}, with the dashed line displaying the acceleration profile at early times, and the solid line corresponding to the present day. \label{f:deltavsims}}
\end{center}
\end{figure}

Figure \ref{f:deltavsims} depicts $\Delta v_{LOS,z}$ over a time baseline of ten years for the simulation of the Antlia 2 dwarf interacting with the Milky Way, and a simulation of the Sgr dwarf interaction (as described in Chakrabarti et al. 2019).  Both simulations use observationally realistic orbits derived from the Gaia proper motions.  The simulation are initialized with a more massive dark matter halo relative to \cite{DOnghiaAguerri2020}, by about a factor of two. The present day acceleration profile is shown in the solid lines, and the acceleration profile at early times (prior to the interaction with the dwarf galaxy) is shown in the dashed line.  The simulation of the Antlia 2 dwarf reproduces the observed planar HI disturbances in the outer disk \citep{LevineBlitzHeiles2006}.  The Antlia 2 dwarf galaxy radial location is close to that of a predicted dwarf galaxy \citep{ChakrabartiBlitz2009} that recently perturbed our Galaxy.  While these simulations of Milky Way-like galaxies recover aspects of the observed Galaxy, such as the observed HI disturbances \citep{ChakrabartiBlitz2009,ChakrabartiBlitz2011}, and large-scale properties of moving groups in the Galactic disk (Craig et al. 2020), we do not resolve the Solar neighborhood.  Therefore, we take the Sun to be along a ring a radius r = 8.1 kpc, and calculate the acceleration along vertical lines of sight at various azimuths.  The solid line shows the average value of $\Delta v_{LOS,z}$ and the errors show the standard deviation.  In contrast to the models shown in Figure \ref{f:static}, both these simulations show a clear asymmetry in the acceleration profile, particularly for $|z| > 1$ kpc relative to the Galactic mid-plane.  Moreover, the acceleration profile develops this asymmetry following the interaction, as is clear from comparing the early-time (i.e., prior to the interaction) acceleration profile (shown in the dashed lines) with the present-day acceleration profile, where the latter is distinctly more asymmetric. 

It is not surprising that the Sgr dwarf interaction shows a more prominent vertical asymmetry, as it is on a polar orbit, relative to the Antlia 2 interaction.  The Antlia 2 dwarf galaxy is on a nearly co-planar orbit and excites large planar disturbances in the Galactic disk, leading to a larger standard deviation at various azimuths, compared to the Sgr interaction where the variation along azimuth is smaller.  An observed asymmetry in the acceleration profile may be the signature of a perturbing dwarf galaxy.  The effects of multiple perturbers in cosmological simulations may lead to more complex vertical acceleration profiles.    

\section{Effects of contaminants}

In order to detect the Galactic acceleration, we have to carefully select our sample of stars.  We select cool stars with low radial velocity "jitter" (e.g. \cite{Wright2005}), such that with repeated $N$ measurements, one may expect to improve our precision by $1/\sqrt{N}$ to the level of 10 cm/s.  Specifically, we select stars from Gaia DR-2 (Gaia DR-2; Gaia Collaboration et al. 2018) that are expected to have low RV jitter on the basis of their stellar parameters (Yu et al. 2018) and their Gaia colors, which recent work has shown can be translated to an empirical constraint on the stellar jitter \citep{Luhn2020gaia}.  As discussed in \citep{Luhn2020gaia}, the metric $\Delta G$ corresponds to evolved stars which correlate with low stellar jitter (for $\Delta G < 1.4$), which can be identified from their Gaia colors, and distance from the main-sequence.  We choose slightly evolved sub-giants as a compromise between selecting stars at the ‘jitter minimum’ and intrinsically bright stars observable at high RV precision at kpc distances.  In addition to stellar jitter, another contaminant is radial velocity variations due to planetary companions and stellar binaries.  Below, we consider in turn the contribution from planetary companions, stellar binaries, and stellar jitter and instrumental noise to the Galactic signal. 

\subsection{Planets \& binaries}

\begin{figure}[ht]        
\begin{center}
\includegraphics[scale=0.48]{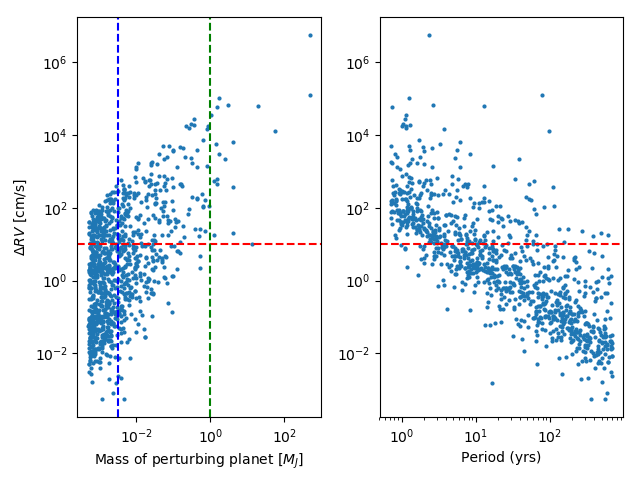}
\caption{$\Delta v_{LOS}$ over a time baseline of ten years (a) as a function of the perturbing planet mass, and (b) as a function of the period, shown for circular Keplerian orbits. The red horizontal line marks $\Delta RV$ = 10 cm/s and the dashed blue and green vertical lines mark Earth and Jupiter mass planets respectively. 
}
\label{f:RVmp}
\end{center}
\end{figure}

\begin{figure}
\begin{center}
\includegraphics[scale=0.48]{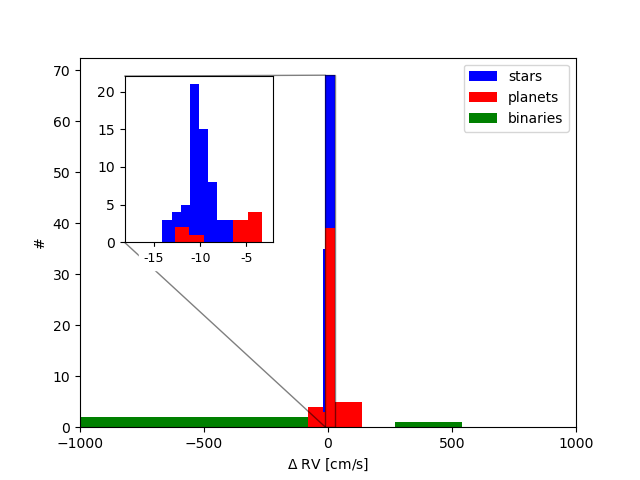}
\caption{Top panel: histogram of $\Delta RV$ over a ten year baseline in a population of single stars, stars with planets on circular Keplerian orbits, and stellar binaries, color coded by the object type (single stars, stars with planets, binaries).  The inset shows a zoom-in centered on the mean of the single star's $\Delta RV$, and out to $\pm$ 5-sigma from the mean.  
\label{f:acc}}
\end{center}
\end{figure}

Figure \ref{f:RVmp} shows $\Delta$ RV as a function of the perturbing planet mass (left panel) and period of orbit (right panel), for a thousand realizations.  This is the parent population of planets that we draw from in constructing the synthetic population we have simulated.  Here, we have considered circular Keplerian orbits with random inclination and phase, with a distribution for the semi-major axis $a$ that follows $\rm log~\Delta a/a = \rm constant$ (in the range of 1-100 AU), motivated by observations 
\citep{Nielsen2019}, and a planet mass distribution that follows $m_{p}^{-2}$, where $m_{p}$ is the planet mass (in the range of Mercury mass to 500 Jupiter masses).  

We have calculated accelerations instantaneously, which is a poor approximation for short period ($<$ 10 year) periods but, only the long-period ($>$ 10 year), low-mass ($10^{-2} M_{J}$) planets contribute to the regime where one would make the Galactic acceleration measurement, i.e., for $\Delta RV \sim $ 10 cm/s.  As we find below, the relative fraction of this contaminant to the Galactic signal is very small. Furthermore, with continuous monitoring, short-period systems can be identified and culled from the sample.  The stellar binary population is similarly drawn from distributions outlined in \cite{Stonkute2018} for sub-giants, assuming circular Keplerian orbits with random inclination and phase; for simplicity we assume the binaries have masses of one and two solar masses.  

To analyze the contribution of planets and binaries to the Galactic acceleration signal, we create a synthetic population composed of single stars that probe the Galactic acceleration, stellar binaries, and stars with planetary companions, adopting observed fractions of stellar binaries for sub-giants ($\sim$ 30 \%) \citep{Stonkute2018}, and planet occurrence rate ($\sim$ 7 \%) \citep{Nielsen2019}.  There is at present no survey that fully encompasses the range of period, semi-major axis, and planet mass that we are interested in.  Therefore, our results are based on an extrapolation of presently observed planet demographics, which mainly probes the high mass end of the planet distribution around sub-giant stars.  We choose a normalization for the planet mass distribution that ensures that observed planet demographics are reproduced (by assumption, 50 \% of the stars in our synthetic population are assigned three planetary companions following the distributions above, which leads to a mean number of about two planets per star).  This typically leads to a few percent of the population having massive planetary companions (with masses between Jupiter mass to 20 times the Jupiter mass and semi-major axis between 10-100 AU), which is consistent with the 1-sigma range of the observed massive planet occurrence rate fraction for sub-giants \citep{Nielsen2019}.
 
To whittle the sample of observable stars down to cool, low jitter stars with metallicity close to solar (such that current instruments can achieve RV precision $\sim$ 1 m/s) at kpc distances that are observable by current-generation instruments to high RV precision, we take the following cuts in height ($|z| > 2$ kpc), temperature ($T_{\rm eff} < 6600~\rm K $), $\Delta G$ ($1.5 > \Delta G > 0.14$), and magnitude (G mag $<$ 15), from Gaia DR-2, which leaves a total number of 124 stars, when we consider a metallicity fraction ($\sim$ 15 \%) of halo stars \citep{Conroy2019} of [Fe]/H] $>$ -0.5.  For simplicity, here we take the Galactic acceleration signal to be a Gaussian, with a mean value equal to the expected signal at a vertical height of $\sim$ 3 kpc ($-3.18 \times 10^{-8} \rm cm/s^{2}$), and a standard deviation equal to 30 \% of this value.  

Figure \ref{f:acc} displays the resultant histogram of $\Delta RVs$ from such a non-homogeneous population, for single stars, stellar binaries, and stars with planets.  To display the contribution from planets to the Galactic signal, the inset shows a zoom-in centered close to the mean of the single stars' $\Delta RVs$ within $\pm$ 5-sigma of the mean, while the larger figure has wider bins to display the full population.  Binaries (shown in green) have large $\Delta$ RV \citep{Stonkute2018}, and thus they separate out from the planet and the Galactic acceleration signal from single stars, and therefore would be easily discarded.  Although stars with long-period, low-mass planets are a contaminant to the Galactic acceleration, their contribution, as shown in the inset, is a small fraction of the signal from single stars (typically about 3 \% within 3-sigma).  The exact contribution varies from realization to realization due to Poisson noise.  Converting from $\Delta RVs$ to accelerations (over a ten year baseline), gives for a typical realization a mean acceleration for single stars of $-3.1 \times 10^{-8} \rm cm/s^{2}$, with a standard deviation of $6.6 \times 10^{-9} \rm cm/s^{2}$.  The mean acceleration of all stars (including the stars with planets) that fall within $\pm$ 5-sigma of the mean of the single stars is $-3.12 \times 10^{-8} \rm cm/s^{2}$, with a 1-sigma of $7.87 \times 10^{-9} \rm cm/s^{2}$.  The p-value from the Kolmogorov-Smirnov test for the two distributions corresponding to all stars that fall within $\pm$ 5-sigma of the mean of the single stars, and stars with planets that fall in this range is consistently $\sim 10^{-4}$ or lower, indicating that these two distributions are clearly distinct. Thus, we can reject the null hypothesis that the signal is due to stars with planetary companions at high confidence.  The mean acceleration experienced by standard RV stars is much less than the mean acceleration for the planet population, which is due to the fact that the majority of stars with planets have at least one with a large acceleration.      

\subsection{Sources of noise in RV data: the LCES HIRES/Keck Precision Radial Velocity Exoplanet Survey}

\begin{figure}[ht]        
\begin{center}
\includegraphics[scale=0.5]{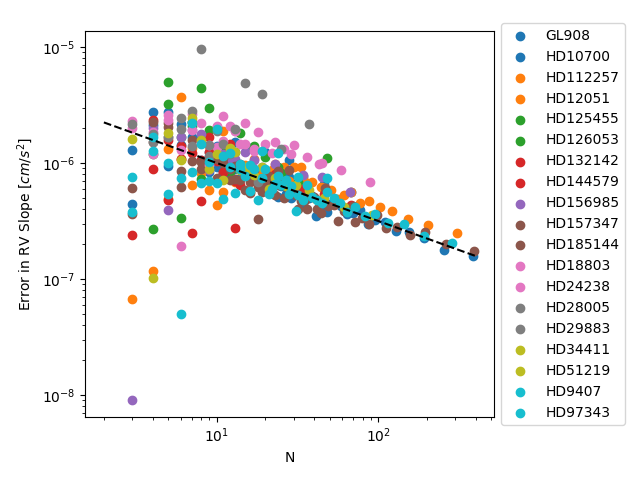}
\caption{Error in the RV slope for the LCES/HIRES Keck observations as a function of sample size, $N$, for standard RV stars from Butler et al. (2017).  Each star is color coded and labeled in the legend. Overlaid is a dashed line that varies as $N^{-1/2}$, where $N$ is the number of observations. \label{f:errors}}
\end{center}
\end{figure}

Observational data will be affected by both stellar "jitter" that arises from intrinsic stellar variability, including stellar oscillations, granulation, short-term activity from stellar rotation and long-term activity due to magnetic fields \citep{Yu2018}, as well as instrumental noise.  The expected contribution to RV jitter for subgiants on day to month timescale from oscillations and granulation is $\sim$ 1.5 m/s \citep{Yu2018}.  These sources of noise (stellar jitter and instrumental noise) are non-Gaussian.  An existing long-term (more than a decade) RV data set of stars near the Sun was produced by the Lick Carnegie Exoplanet Survey Team (LCES), using radial velocities from HIRES on Keck, and is described in \cite{Butler2017}.  To determine if we may model observational sources of noise as being effectively Gaussian for the purposes of the acceleration measurement, we select standard RV stars from the data from \cite{Butler2017}, and calculate the error in the slope of the RV curve.  We only consider data from June 2004 onwards to avoid the discontinuity in the data, which still yields nearly ten years of RV data.  The typical RV precision of this dataset is $\sim$ 1 m/s. 

Figure \ref{f:errors} shows the error in the RV slope for 18 standard RV stars as we select smaller samples of the total observational sample for a given star (removing every n'th observation, from n=2, up to half the observations, while holding the baseline constant).  As is clear, except for small sample sizes ($<$ 20), the error in the slope scales as $1/\sqrt(N)$, where $N$ is the sample size.  Therefore, we may reasonably expect that $N$ independent observations of the same star will serve to effectively increase the RV precision of observations as $1/\sqrt{N}$.  For a long-term monitoring survey, one may then carry out individual RV measurements at some threshold precision, for example, $\sim $ 1 m/s RV precision for individual measurements, and thus measure the acceleration with precisions approaching $\sim$ 10 cm/s over a baseline of ten years from a hundred independent measurements. There are hopes and expectations that the problem of RV jitter can be mitigated or solved and that the measurement uncertainty of center-of-mass motions of stars can be reduced towards the instrumental precision, further improving the precision of the acceleration measurements.  These methods will rely on simultaneously measured activity indices, which are collected as part of the measurement (e.g. Ca lines). 

The average slope of the RV curve for this sample of 18  standard RV stars is $-5.2 \times 10^{-8} \rm cm/s^{2} \pm 7 \pm 10^{-7} \rm cm/s^{2}$.  Due to the large error here, we cannot determine the local Galactic acceleration accurately, but the average value is nevertheless consistent within the errors with expectations of the local Galactic acceleration based on models for stars within a few hundred pc of the Sun.  

\section{Conclusions}

We have analyzed the vertical acceleration experienced by stars in the Galaxy in the context of a number of different models of the Milky Way:  static models, isolated high-resolution simulations, and simulations that include interactions between the Milky Way and dwarf galaxies.  The magnitude of the change in the line-of-sight velocity over ten year baselines is $\sim$ few - 10 cm/s at kpc distances off the Galactic mid-plane in static models (with a significant dependence on the mass of the dark matter halo), and in isolated simulations.  Simulations of the Milky Way interacting with dwarf galaxies have distinctly asymmetric vertical accelerations, especially for $|z| > 1$ kpc relative to the Galactic mid-plane.  We find that although  low-mass ($< 10^{-2} M_{\rm Jupiter}$), long-period ($>$ 10 year) planets are a contaminant, they do not overwhelm the Galactic acceleration signal in a realistic sample size of stars selected from Gaia DR-2 that are currently observable.  We find that one can reject the null hypothesis that the signal is due to stars with planetary companions at high confidence. 

We have analyzed ten-year data of standard RV stars from the LCES HIRES/Keck precision radial velocity survey.  Although the error in the RV slope is non-Gaussian, we find that we may consider it to be effectively Gaussian for the purposes of the Galactic acceleration measurement, as long as a sufficient number of epochs are obtained ($N > 20$).  High precision RV measurements at kpc distances will enable us to determine the total density directly from the Poisson equation, and the dark matter density given an accounting of the baryon budget; they will also provide a direct view of dark matter sub-structure in the Milky Way.

Analyzing cosmological simulations with both cold-dark matter models as well as alternatives to cold dark matter, such as self-interacting dark matter \citep{SpergelSteinhardt2000,TulinYu2018} will help us understand if the nature of the dark matter particle produces measurable differences at these scales.  Determining the Galactic acceleration lies at the nexus of three areas that are often disparate -- dark matter detection, studies of planet demographics, and Galactic dynamics, and can potentially produce discoveries in all three areas.\\

\bigskip
\bigskip
\bigskip 

\acknowledgments
SC acknowledges support from NASA ATP NNX17AK90G,
NSF AAG grant 1517488, and from Research Corporation
for Scientific Advancement's Time Domain Astrophysics
Scialog. KLR acknowledges support from NSF AAG grant 1615483.

The Center for Exoplanets and Habitable Worlds and the Penn State Extraterrestrial Intelligence Center are supported by the Pennsylvania State University and the Eberly College of Science.

%\newpage
\bibliographystyle{aasjournal}
\bibliography{bibl}

\end{document}